\documentclass[aps,floats,amssymb,amsmath,prb,onecolumn]{revtex4}
\usepackage{calc}
\usepackage{psfrag}
\usepackage{graphicx}
\usepackage{color}

\begin{document}
\title{$p$-wave superconductivity on monolayer and bilayer honeycomb lattice}

\author{M. V. Milovanovi\'c}
\affiliation{Scientific Computing Laboratory, Institute of Physics
Belgrade,
 University of Belgrade, Pregrevica 118, 11 080 Belgrade, Serbia}

\begin{abstract}
We derive ground state wave functions of superconducting
instabilities on the honeycomb lattice induced by nearest-neighbor
attractive interactions. They reflect the Dirac nature of electrons
in the low-energy limit. For the order parameter that is the same
irrespective of the direction to any of the nearest neighbors we
find weak pairing (slowly decaying) behavior  in the orbital part of
the Cooper pair with no angular dependence. At the neutrality point,
in the spin-singlet case, we recover a strong pairing behavior. We
also derive ground state wave functions for the superconductivity on
the bilayer honeycomb lattice, with strong interlayer coupling,
induced by attractive interactions between sites that participate in
a low-energy description. Without these interactions, free electrons
are described by a Dirac equation with a quadratic dispersion. This
unusual feature, similarly to $^3$He - B phase, leads to the
description with two kinds of Cooper pairs, with $p_x + i p_y$ and
$p_x - i p_y$ pairing, in the presence of the attractive
interactions. We discuss the edge modes of such a spin-singlet
superconductor and find that it represents a trivial topological
superconductor.
% Due to the spin and valley degrees of freedom we
%expect doubling of Majorana modes on the edge of spin-singlet
%superconductor i.e. it represents a trivial topological
%superconductor.
\end{abstract}
\maketitle
\section{Introduction: Superconductivity on honeycomb lattice}
The advent of graphene \cite{disc} opened a door for exploration of
new phenomena in two-dimensional Dirac-like condensed matter
systems. One of the intriguing questions is of superconducting
correlations of electrons on the honeycomb lattice system.
Superconductivity has been induced in short graphene samples through
proximity effect with superconducting contacts \cite{hee}. This
indicates  that Cooper pairs can propagate coherently in graphene.
In principle superconductivity on the graphene honeycomb lattice can
be induced by short-range attractive interactions and explorations
of allowed possibilities were given in Refs. \cite{ucho,abs,herr}.
Among the most interesting is the so-called $p + i p$
superconducting instability introduced in Ref. \cite{ucho}. It would
be supported by the most natural nearest-neighbor attractive
interaction and have distinct features of the Dirac electrons. Later
it was showed \cite{herr}, by a restricted (low-energy) analysis,
that this state may be less energetically favorable  with respect to
Kekule-like order parameter arrangements.  Nevertheless, the $p + i
p$ instability seems, though an exotic state, a very attractive
possibility because of its underlying symmetry of the order
parameter, the same as for Pfaffian quantum Hall state \cite{mr} or
$p + i p$ spinless superconductor \cite{rg}. The later systems
support non-Abelian statistics, which is at the heart of the idea of
the topological computing \cite{rmp}. There is an important
difference between these states and the proposed graphene state. The
superconducting instability in graphene does not break time-reversal
symmetry  and those systems do. Due to the valley degeneracy we
effectively have two $(p \pm i p)$ order parameters and that
requires additional understanding of intertwined correlations and
underlying symmetries. One way, just as in the Pfaffian state
\cite{rg}, is to look for the ground state wave function and
recognize the structures and symmetries.

In this paper, in the first part, we will find the effective
(long-distance) expression for the ground state wave function of the
$p + i p$ spin-singlet instability described in Ref.
\onlinecite{ucho} and display pertinent symmetries in this case.
Also a spinless case will be discussed. We will use the BCS
mean-field formalism. In the following section we will set up the
BCS formalism, solve the Bogoliubov - de Gennes (BdG) equations and
find the expression for the ground state wave functions. The last
section of the first part is devoted to conclusions. The second part
of the paper is devoted to the $p \pm i p$ superconductivity on the
bilayer honeycomb lattice. We refer reader to this part of the paper
for an introduction.

\section{Superconductivity on honeycomb lattice and its ground states}

The Hamiltonian for free electrons on the honeycomb lattice is
\begin{equation}
H_{0} = - t \sum_{\langle i j \rangle} \sum_{\sigma = \uparrow,
\downarrow}(a_{i, \sigma}^{\dagger} b_{j, \sigma} + h.c.) - \mu
\sum_{i} \hat{n}_{i},\label{ho}
\end{equation}
where $t$ is the hopping energy between nearest neighbor C (carbon)
atoms, $a_{i, \sigma} (a_{i, \sigma}^{\dagger})$ is the on-site
annihilation (creation) operator for electrons in the sublattice A
with spin $\sigma = \uparrow, \downarrow$, and $b_{i, \sigma} (b_{i,
\sigma}^{\dagger})$ for sublattice B, $\hat{n}_{i}$ is the on-site
number operator, and $\mu$ is the graphene chemical potential. We
use units such that $\hbar = 1$. Diagonalization of Eq.(\ref{ho})
leads to a spectrum given by: $\epsilon_{\vec{k}} = \pm t
|S(\vec{k})|$, where $\vec{k}$ is the two-dimensional momentum, and
$S(\vec{k}) = \sum_{\vec{\delta}} \exp\{i \vec{k} \vec{\delta}\}$
with $\vec{\delta}$'s defined as $\delta_{1} = a (0, 1/\sqrt{3})$,
$\delta_{2} = a/2 (1, - 1/\sqrt{3})$, and $\delta_{3} = a/2 (- 1, -
1/\sqrt{3})$, and $ a = \sqrt{3}\; a_{cc}$, $a_{cc}$ is the distance
between $C$ atoms and $a$ is the next to nearest neighbor distance.
At the corners of the hexagonal Brillouin zone, $\vec{K}_{\pm} = (2
\pi)/a (\pm 2/3,0)$, we have $S(\vec{K}_{\pm} + \vec{k}) \approx \mp
a \sqrt{3}/2 (k_{x} \mp i k_{y})$, and the band has the shape of a
Dirac cone: $\epsilon(\vec{K}_{\pm} + \vec{k}) = \pm v_{F}
|\vec{k}|$, where $v_F = (\sqrt{3} a t)/2 $ is the Fermi-Dirac
velocity.

For the sake of simplicity we will consider only nearest-neighbor
attractive interactions among electrons. The on-site repulsive
interactions can be introduced and will not change our conclusions.
Therefore the complete Hamiltonian will include nearest-neighbor
interactions as follows,
\begin{equation}
H_{I} = g \sum_{\langle i j \rangle} \sum_{\sigma, \sigma^{'}} a_{i,
\sigma}^{\dagger} a_{i, \sigma} b_{j, \sigma^{'}}^{\dagger} b_{j,
\sigma^{'}},
\end{equation}
where $ g < 0$. We will assume the spin-singlet pairing among
nearest-neighbors and apply the BCS ansatz with $ \Delta_{ij} =
\langle a_{i, \downarrow} b_{j, \uparrow} - a_{i, \uparrow} b_{j,
\downarrow} \rangle$, the superconducting order parameter.
Furthermore we assume one and the same $\Delta_{ij} = \Delta$ for
all nearest neighbors, which due to global gauge $(U(1))$
transformations on $a$'s and $b$'s can be chosen real and positive
\cite{polet}. The interaction part, $H_{I}$, becomes
\begin{equation}
\tilde{H}_{BCS} = \{ g \sum_{\langle i j \rangle} \Delta ( a_{i,
\uparrow}^{\dagger} b_{j, \downarrow}^{\dagger} - a_{i,
\downarrow}^{\dagger} b_{j, \uparrow}^{\dagger}) + h.c.\} - 3 g
|\Delta|^{2}.
\end{equation}
The order parameter in the momentum space is
\begin{equation}
\Delta_{\vec{k}} = \sum_{\langle i j \rangle} \Delta \exp\{i \vec{k}
(\vec{i} - \vec{j})\} = \Delta \sum_{\vec{\delta}} \exp\{i \vec{k}
\vec{\delta}\} = \Delta S(\vec{k})
\end{equation}
Therefore near $K$ points $\Delta_{\vec{K}_{\pm} + \vec{k}} \sim \mp
 (k_{x} \mp i k_{y})$, which then describes two $p$-wave like
 superconducting order parameters in a low effective description.
 The complete BCS Hamiltonian can be now cast in the following form
 in the momentum space,
 \begin{equation}
 H_{BCS} = \sum_{\vec{k}} \phi_{\vec{k}}^{\dagger} M_{\vec{k}}
 \phi_{\vec{k}}, \label{hbcs}
 \end{equation}
 where
 \begin{equation}
 \phi_{\vec{k}}^{\dagger} = (a_{\vec{k} \uparrow}^{\dagger}, b_{\vec{k} \uparrow}^{\dagger}
a_{-\vec{k} \downarrow}, b_{-\vec{k} \downarrow})
\end{equation}
with defined $ a_{\vec{k} \sigma} = \sum_{i} a_{i \sigma} \exp\{ i
\vec{k} \; \vec{i}\}$ and $ b_{\vec{k} \sigma} = \sum_{i} b_{i
\sigma} \exp\{  i \vec{k} \; \vec{i}\}$, and, with $ g \Delta \equiv
\Delta$ for short,
\begin{displaymath}
M_{\vec{k}} = \left[\begin{array}{cccc} - \mu & - t S(\vec{k}) & 0 & \Delta S(\vec{k}) \\
                                    - t S^{*}(\vec{k})& - \mu & \Delta S(-\vec{k})  & 0 \\
                          0 & \Delta S^{*}(-\vec{k}) & \mu &  t S(\vec{k})  \\
                          \Delta S^{*}(\vec{k}) & 0 & t S^{*}(\vec{k})  &
                          \mu
\end{array} \right].
\end{displaymath}
We look for the solution in the form of a diagonalized Bogoliubov
BCS Hamiltonian,
\begin{equation}
H_{BCS} = \sum_{\vec{k}, \gamma = \pm} \omega_{\vec{k},
\gamma}^{\alpha} \alpha^{\dagger}_{\vec{k}, \gamma} \alpha_{\vec{k},
\gamma} + \sum_{\vec{k}, \gamma = \pm} \omega_{\vec{k},
\gamma}^{\beta} \beta^{\dagger}_{\vec{k}, \gamma}  \beta_{\vec{k},
\gamma} + E_{0}\;,
\end{equation}
where $\alpha_{\vec{k}, \gamma}$ and $\beta_{\vec{k}, \gamma}$,
$\gamma =\pm $ are new quasiparticles at momentum $\vec{k}$. For the
dispersions we have:
\begin{equation}
\omega_{\vec{k}, \gamma}^{\alpha} = \gamma
\omega_{\vec{k}}^{\alpha}\;\;\; {\rm and}\;\;\; \omega_{\vec{k},
\gamma}^{\beta} = \gamma \omega_{\vec{k}}^{\beta}\;,
\end{equation}
where $\gamma = \pm $. We define a general solution $\alpha$ as
\begin{equation}
\alpha_{\vec{k}} = u_{\vec{k}, \uparrow} a_{\vec{k}, \uparrow} +
v_{\vec{k}, \uparrow} b_{\vec{k}, \uparrow} + u_{\vec{k},
\downarrow} a_{- \vec{k}, \downarrow}^{\dagger} + v_{\vec{k},
\downarrow} b_{- \vec{k}, \downarrow}^{\dagger}.
\end{equation}
Next we have to solve the Bogoliubov - de Gennes (BdG) equations,
which follow from the following condition,
\begin{equation}
 [\alpha_{\vec{k}}, H_{BCS}] = E \alpha_{\vec{k}}.\label{beq}
\end{equation}
From this matrix eigenvalue problem we obtain energies of the
Bogoliubov quasiparticles,
\begin{equation}
E_{p} = \pm \sqrt{(v_F |S(\vec{k})| + p \mu)^{2} + |\Delta
S(\vec{k})|^{2}},
\end{equation}
where $\pm$ stands for the particle and hole branches respectively
for two kinds of excitations $p = - 1 (\alpha)$ and $ p = + 1
(\beta)$. For $\mu = 0$ the system is gapless and we need a coupling
$g$ larger than a critical value for the superconducting instability
to exist \cite{ucho}. This can be found considering in the BCS
formalism the consistency or gap equation.

For each valley we have to solve the Bogoliubov problem using the
expansion $S(\vec{K}_{\pm} + \vec{k}) \approx \mp a \sqrt{3}/2
(k_{x} \mp i k_{y})$. Near $K_+$ we need to diagonalize the
following matrix, $M^{*}_{\vec{k}}$, that comes out of
Eq.(\ref{beq}):
\begin{displaymath}
\left[\begin{array}{cccc} - \mu & v_{F} k & 0 & s k \\
                                    v_{F} k^{*} & - \mu & s k^{*}  & 0 \\
                          0 & s  k & \mu & - v_{F} k  \\
                          s  k^{*} & 0 & - v_{F} k^{*}  &
                          \mu
\end{array} \right],
\end{displaymath}
where $ s = s^{*} = - \Delta a \sqrt{3}/2 > 0$. Its eigenvectors
(after normalization) enter the following expressions for Bogoliubov
quasiparticles:
\begin{equation}
\alpha_{\vec{k},+} = \frac{1}{2 \sqrt{E_{\alpha}[E_{\alpha} - (\mu -
v_F |k|)]}} \{[E_{\alpha} - (\mu - v_F k)] (\sqrt{\frac{k}{k^{*}}}
a_{+ \uparrow} + b_{+ \uparrow}) + s |k| (\sqrt{\frac{k}{k^{*}}}
a_{- \downarrow}^{\dagger} + b_{- \downarrow}^{\dagger})\},
\end{equation}
and
\begin{equation}
\beta_{\vec{k},+} = \frac{1}{2 \sqrt{E_{\beta}[E_{\beta} - (\mu +
v_F |k|)]}} \{[E_{\beta} - (\mu + v_F k)] (\sqrt{\frac{k}{k^{*}}}
a_{+ \uparrow} - b_{+ \uparrow}) - s |k| (\sqrt{\frac{k}{k^{*}}}
a_{- \downarrow}^{\dagger} - b_{- \downarrow}^{\dagger})\},
\label{beta1}
\end{equation}
and quasiholes:
\begin{equation}
\alpha_{\vec{k},-} = \frac{1}{2 \sqrt{E_{\alpha}[E_{\alpha} + (\mu -
v_F |k|)]}} \{-[E_{\alpha} + (\mu - v_F k)] (\sqrt{\frac{k}{k^{*}}}
a_{+ \uparrow} + b_{+ \uparrow}) + s |k| (\sqrt{\frac{k}{k^{*}}}
a_{- \downarrow}^{\dagger} + b_{- \downarrow}^{\dagger})\},
\end{equation}
and
\begin{equation}
\beta_{\vec{k},-} = \frac{1}{2 \sqrt{E_{\beta}[E_{\beta} + (\mu +
v_F |k|)]}} \{-[E_{\beta} + (\mu + v_F k)] (\sqrt{\frac{k}{k^{*}}}
a_{+ \uparrow} - b_{+ \uparrow}) - s |k| (\sqrt{\frac{k}{k^{*}}}
a_{- \downarrow}^{\dagger} - b_{-
\downarrow}^{\dagger})\},\label{beta2}
\end{equation}
for the Bogoliubov solution near point $\vec{K}_+$, where we denoted
$a_{\vec{K}_\pm \pm \vec{k}, \sigma} \equiv a_{\pm \sigma}$ and
$b_{\vec{K}_\pm \pm \vec{k}, \sigma} \equiv b_{\pm \sigma}$.

The natural eigenstates of chirality appeared in our expressions.
For example $(\sqrt{\frac{k}{k^{*}}} a_{+ \uparrow} + b_{+
\uparrow})$ represents spinor:
\begin{equation}
\chi = \left[\begin{array}{c}  \sqrt{\frac{k^{*}}{k}} \\
                                   1
\end{array} \right],\label{spinor}
\end{equation}
which is the eigenstate of the chirality operator $
\frac{\vec{\sigma} \vec{k}}{|k|}$, defined with $\vec{\sigma} =
(\sigma_{x},\sigma_{y})$ Pauli matrices, i.e. the pseudospin (due to
two sublattices) is along the momentum vector. The state
$(\sqrt{\frac{k^{*}}{k}} a_{- \downarrow} + b_{- \downarrow})$
represents the same spinor because of the interchanged roles of
sublattices in the $\vec{K}_{-}$ point. To see this in more details
we would like to remind the reader that instead of  the Dirac free
electron representation by the spinor
\begin{equation}
 \chi_{\vec{k}}^{\dagger} = (a_{\vec{K}_{+} + \vec{k}, \sigma}^{\dagger}, b_{\vec{K}_{+} + \vec{k}, \sigma}^{\dagger}
b_{\vec{K}_{-} + \vec{k}, \sigma}^{\dagger}, a_{\vec{K}_{-} +
\vec{k}, \sigma}^{\dagger}),
\end{equation}
and the chirality operator is defined as
\begin{equation}
\left[\begin{array}{cc} \frac{\vec{\sigma} \vec{k}}{|\vec{k}|} & 0 \\
                        0 &  - \frac{\vec{\sigma} \vec{k}}{|\vec{k}|}
\end{array} \right],\label{chiop}
\end{equation}
in the BdG formalism we work with
\begin{eqnarray}
 \phi_{\vec{k}}^{\dagger}& = &(a_{\vec{K}_{+} + \vec{k}
\uparrow}^{\dagger}, b_{\vec{K}_{+} + \vec{k} \uparrow}^{\dagger}
a_{\vec{K}_{-} - \vec{k} \downarrow}, b_{\vec{K}_{-} - \vec{k}
\downarrow}) \nonumber \\
&\equiv&(a_{+ \uparrow}^{\dagger}, b_{+ \uparrow}^{\dagger} a_{-
\downarrow}, b_{- \downarrow}).
\end{eqnarray}
Note the reversed order of sublattices and the change of the sign of
the momentum $\vec{k}$ near $\vec{K}_{-}$ point in the BdG formalism
with respect to the free one. Thus the lower $2 \times 2$ matrix on
the diagonal of the Hamiltonian matrix in the free Dirac case can be
read off from:
\begin{equation}
\left[\begin{array}{cc}  b_{\vec{K}_{-} + \vec{k},
\sigma}^{\dagger}& a_{\vec{K}_{-} + \vec{k}, \sigma}^{\dagger}
\end{array} \right]
\left[\begin{array}{cc} - \mu & - v_F k^* \\
                        - v_F k &  - \mu
\end{array} \right]
\left[\begin{array}{c}  b_{\vec{K}_{-} + \vec{k}, \sigma}\\
a_{\vec{K}_{-} + \vec{k}, \sigma}
\end{array} \right],\label{Drep}
\end{equation}
i.e. it is equal to $ - v_F \vec{k} \vec{\sigma} - \mu$. Note that
if we change the sign of $\vec{k}$ vector in Eq.(\ref{Drep}) i.e.
$\vec{k} \rightarrow - \vec{k}$ the off-diagonal elements in the
matrix will change the sign, so that in this basis in the free
representation the chirality operator will not have minus sign in
the lower right entry of the matrix representation in
Eq.(\ref{chiop}). Therefore $(\sqrt{\frac{k^{*}}{k}} a_{-
\downarrow} + b_{- \downarrow})$ represents the same spinor (up to a
phase factor) as in Eq.(\ref{spinor}) and the same chirality
eigenstate (with positive eigenvalue) as we pointed out earlier.
Nevertheless in the Bogoliubov representation we still have
\begin{equation}
\left[\begin{array}{cc}  a_{- \downarrow}& b_{- \downarrow}
\end{array} \right]
\left[\begin{array}{cc} \mu & - v_F k^* \\
                        - v_F k &   \mu
\end{array} \right]
\left[\begin{array}{c}  a_{- \downarrow}^{+}\\
b_{- \downarrow}^{+}
\end{array} \right],
\end{equation}
i.e. the matrix is $- v_F \vec{k} \vec{\sigma} + \mu$, and the
representation of the chirality operator stays the same as in
Eq.(\ref{chiop}). We will use this fact later.
 On the other hand the combinations in Eqs. (\ref{beta1}) and
 (\ref{beta2}):
$(\sqrt{\frac{k}{k^{*}}} a_{+ \uparrow} - b_{+ \uparrow})$ and
$(\sqrt{\frac{k^{*}}{k}} a_{- \downarrow} - b_{- \downarrow})$ have
the pseudospin vector in the opposite direction of the momentum
vector $\vec{k}$.

It is thus natural to introduce the following notation:
\begin{eqnarray}
\sqrt{\frac{k}{k^{*}}} a_{+ \uparrow} + b_{+ \uparrow} \equiv c_{+
\uparrow v},\\
\sqrt{\frac{k}{k^{*}}} a_{- \downarrow}^{\dagger} + b_{-
\downarrow}^{\dagger}\equiv c_{- \downarrow v}^{\dagger},\\
\sqrt{\frac{k}{k^{*}}} a_{+ \uparrow} - b_{+ \uparrow} \equiv c_{+
\uparrow w},\\
-\sqrt{\frac{k}{k^{*}}} a_{- \downarrow}^{\dagger} + b_{-
\downarrow}^{\dagger}\equiv c_{- \downarrow w}^{\dagger},
\end{eqnarray}
where $v$ and $w$ denote the chirality i.e. whether the pseudospin
vector is along or in the opposite direction with respect to the
$\vec{k}$ vector, respectively. We have to note that these electron
operators are defined up to a phase factor, most importantly
$\sqrt{\frac{k}{k^{*}}}$ phase. This degree of freedom should not
influence the physics, but we chose the definitions so that  later
the symmetry under exchange of particles in the ground state wave
function is transparent.

The $\alpha$ and $\beta$ sectors are obviously decoupled in the
Bogoliubov description and we can concentrate and closely examine
the $\alpha$ sector first. Furthermore we do not have to consider
$\vec{K}_{-}$ point separately as the symmetry considerations tell
us that the BdG equations around this point will induce  the
coupling or states of an electron around $\vec{K}_{+}$ point with
$\downarrow$ projection of spin and those around  $\vec{K}_{-}$
point with $\uparrow$ projection of spin.

Thus it suffices to consider $\alpha$ sector first (with $ c_{+
\uparrow v}$ and $ c_{- \downarrow v}$) and then use the symmetry
arguments, more precisely antisymmetry under real spin exchange to
recover the whole ground state wave function. We can rewrite
$\alpha$'s in the following form,
\begin{eqnarray}
\alpha_{k, +} = u^{p}_{k} c_{+ \uparrow v} + v^{p}_{k}
c^{\dagger}_{- \downarrow v} \\
\alpha_{k, -} = u^{h}_{k} c_{+ \uparrow v} + v^{h}_{k}
c^{\dagger}_{- \downarrow v}.
\end{eqnarray}
We should demand $ \alpha_{k, +} |G\rangle = 0$ and $ \alpha_{k,
-}^{\dagger} |G\rangle = 0$, for any $k$, if $|G\rangle$ is to
represent the ground state vector. That implies that in the $\alpha$
sector of $\vec{K}_{+}$ point we have the following contribution to
the ground state,
\begin{equation}
\prod_{k} (u_{k}^{p} - v_{k}^{p} c_{+ \uparrow v}^{\dagger} c_{-
\downarrow v}^{\dagger})|0\rangle,
\end{equation}
where $|0\rangle$ denotes the vacuum. This state is annihilated with
both, $\alpha_{k,+}$ and $\alpha_{k,-}^{\dagger}.$ The symmetry
arguments demand that we should get a similar expression considering
BdG equations at $\vec{K}_{-}$ point. If we denote by $g_{\alpha}(k)
= - \frac{v_{k}^{p}}{u_{k}^{p}}$, the ground state vector in the
$\alpha$ sector should look like:
\begin{eqnarray}
&&\prod_{k} (1 + g_{\alpha}(k) c_{+ \uparrow v}^{\dagger} c_{-
\downarrow v}^{\dagger}) (1 + g_{\alpha}(k) c_{- \uparrow
v}^{\dagger} c_{+ \downarrow v}^{\dagger})|0\rangle\nonumber \\
=&& \prod_{k}\{1 + g_{\alpha}(k) [c_{+ \uparrow v}^{\dagger} c_{-
\downarrow v}^{\dagger} + c_{- \uparrow v}^{\dagger} c_{+ \downarrow
v}^{\dagger}] + \frac{g_{\alpha}^{2}(k)}{2} [c_{+ \uparrow
v}^{\dagger} c_{- \downarrow v}^{\dagger} + c_{- \uparrow
v}^{\dagger} c_{+ \downarrow v}^{\dagger}]^{2}\}|0\rangle \nonumber \\
=&& \exp\{ \sum_{k} g_{\alpha}(k) [c_{+ \uparrow v}^{\dagger} c_{-
\downarrow v}^{\dagger} + c_{- \uparrow v}^{\dagger} c_{+ \downarrow
v}^{\dagger}]\}|0\rangle
\end{eqnarray}
Now we can identify $g_{\alpha}(k)$ to represent a Fourier transform
of the wave function of a Cooper pair of electrons, which is a
spin-singlet with respect to spin degree of freedom and a triplet
state (symmetric under exchange) with respect to valley $(K_{\pm})$
degree of freedom. If we defined differently our electron operators
there would be possibility for $g_{\alpha}(k)$ to acquire the phase
factor $\sqrt{\frac{k}{k^{*}}}$, which would make the identification
of the antisymmetry under exchange harder.

Taking into account the $\beta$ sector (with the chirality in the
opposite direction of the momentum: $w$) the complete ground state
vector is
\begin{equation}
\exp\{ \sum_{k} g_{\alpha}(k) [c_{+ \uparrow v}^{\dagger} c_{-
\downarrow v}^{\dagger} + c_{- \uparrow v}^{\dagger} c_{+ \downarrow
v}^{\dagger}] + \sum_{k} g_{\beta}(k) [c_{+ \uparrow w}^{\dagger}
c_{- \downarrow w}^{\dagger} + c_{- \uparrow w}^{\dagger} c_{+
\downarrow w}^{\dagger}]\}|0\rangle, \label{gsv}
\end{equation}
where
\begin{equation}
g_{\alpha}(k) = - \frac{s |k|}{E_{\alpha} - (\mu - v_F
|k|)}\;\;\;{\rm and}\;\;\; g_{\beta}(k) = - \frac{s |k|}{E_{\alpha}
- (\mu + v_F |k|)}.
\end{equation}
Using the long-distance (low-momentum) expansions for $E_{\alpha}$
and $E_{\beta}$, for finite $\mu$,
\begin{equation}
E_{\alpha(\beta)} \approx \mu \mp v_F |k| + \frac{s^{2} |k|^{2}}{2
\mu},
\end{equation}
we find the long-distance behavior of the pair wave function to be
\begin{equation}
\lim_{|\vec{r}| \rightarrow \infty } g_{\alpha}(\vec{r}) =
\lim_{|\vec{r}| \rightarrow \infty } g_{\beta}(\vec{r}) \sim
\frac{1}{|\vec{r}|},
\end{equation}
i.e. we have a case for a weak pairing \cite{rg}. As emphasized in
Ref. \onlinecite{rg} the term weak pairing does not mean also weak
coupling, it stands for a phase with an unusual large spread of the
Cooper pairs. On the other hand for $\mu = 0$ we have that
$g_{\alpha}(k)$ and $g_{\beta}(k)$ are two constants and the Cooper
pairs are localized on  a short scale $\sim a$ in the graphene
system at the neutrality point. Thus for $\mu = 0$ we have a case
for a strong pairing.

The ground state vector (wave function) in Eq.(\ref{gsv}) displays
two kinds of Cooper pairs, each antisymmetric under combined
exchange of (a) orbital, (b) valley ($\vec{K}_{\pm}$), and (c) spin
$(\uparrow, \downarrow)$ degree of freedom. Two kinds of Cooper
pairs stem from the chirality (sublattice) degree of freedom
intimately connected with the Dirac-nature of the electron with
both, particles and holes. They both, particles (with positive
chirality $v$ at $\vec{K}_{+}$) and holes (with negative chirality
$w$ at $\vec{K}_{+}$), constitute Cooper pairs, which are symmetric
under $v \leftrightarrow w, v_{F} \rightarrow - v_{F}$
transformation.

In the long distance limit we recover the form of the wave function
of ordinary $s$-wave superconductor as given in Ref.
\onlinecite{sch}, though with more, two-component, degrees of
freedom. The Cooper pair wave function is antisymmetric under spin
exchange and symmetric under exchange of valley ($\vec{K}_{\pm}$),
sublattice $(v, w)$, and orbital degrees of freedom.

Next we will discuss the spin-triplet case, more precisely we will
assume that the system is spin-polarized and not consider spin in
the following. Therefore fermions are spinless just like in the
Pfaffian case, but they live on the honeycomb lattice. We will
assume $\langle a_{i} b_{j} \rangle = \Delta$. In this case the
Bogoliubov problem in Eq.(\ref{hbcs}) for the spin-singlet pairing
transforms into a similar one with $a_{\vec{k},\sigma} \equiv
a_{\vec{k}}$ and $b_{\vec{k},\sigma} \equiv b_{\vec{k}}$, and the
matrix $M_{\vec{k}}$ becomes as follows
\begin{displaymath}
M_{\vec{k}} = \left[\begin{array}{cccc} - \mu & - t S(\vec{k}) & 0 & \Delta S(\vec{k}) \\
                                    - t S^{*}(\vec{k})& - \mu & -\Delta S(-\vec{k})  & 0 \\
                          0 & -\Delta S^{*}(-\vec{k}) & \mu &  t S(\vec{k})  \\
                          \Delta S^{*}(\vec{k}) & 0 & t S^{*}(\vec{k})  &
                          \mu
\end{array} \right].
\end{displaymath}
Around the $\vec{K}_{+}$ point we have
\begin{displaymath}
\left[\begin{array}{cccc} - \mu & v_{F} k^{*} & 0 & s k^{*} \\
                                    v_{F} k & - \mu & - s k  & 0 \\
                          0 & - s  k^{*} & \mu & - v_{F} k^{*}  \\
                          s  k & 0 & - v_{F} k  &
                          \mu
\end{array} \right],
\end{displaymath}
where $s = - \Delta a \frac{\sqrt{3}}{2} > 0$ as before. The problem
around the $\vec{K}_{-}$ point is a copy of the problem around the
$\vec{K}_{+}$ point.

Now the $M_{\vec{k}}$ matrix around $\vec{K}_{+}$ point cannot be
cast, as in the spin-singlet case, in the following form,
\begin{displaymath}
\left[\begin{array}{cc} v_F \vec{\sigma} \vec{k} - \mu I_{2} & s \vec{\sigma} \vec{k} \\
                        s \vec{\sigma} \vec{k} &  - v_F \vec{\sigma}
                        \vec{k} + \mu I_2
\end{array} \right],
\end{displaymath}
where $I_2$ is the $ 2 \times 2$ identity matrix, which commutes
with the chirality matrix (Eq.{\ref{chiop}}).
 $M_{\vec{k}}$ around $\vec{K}_{+}$ point can be compactly written
as
\begin{displaymath}
\left[\begin{array}{cc} v_F \vec{\sigma} \vec{k} - \mu I_{2} & s i \vec{k} \times \vec{\sigma}  \\
                        - s i \vec{k} \times \vec{\sigma}  &  - v_F \vec{\sigma}
                        \vec{k} + \mu I_2
\end{array} \right],
\end{displaymath}
and it does not commute with the chirality operator. The eigenstates
of the Bogoliubov problem do not have to be the eigenstates of
chirality. We find the following eigenvalues $E_{p} = \pm
\sqrt{\mu^2 + |\vec{k}|^{2} s^2 + |\vec{k}|^{2} v_F^2 + p\; 2
\sqrt{\mu^2 v_F^2 |\vec{k}|^{2} + s^2 v_F^2 |\vec{k}|^{2}}}$, where
$p = + 1 (\alpha)$ and $ p = - 1 (\beta)$ are two branches as
before. The associated eigenvectors can be written as sums of
fermionic particle eigenstates of chirality only in the low-momentum
limit and we list those connected with positive eigenvalues,
\begin{equation}
\alpha_{\vec{k}, +} = \frac{1}{\sqrt{2(1 + \frac{|k|^2 s^2}{4
\mu^2})}} [\frac{|k| s}{2 \mu} (- \sqrt{\frac{k}{k^*}} a_{+} +
b_{+}) + ( \sqrt{\frac{k}{k^*}} a_{-}^{\dagger} + b_{-}^{\dagger})],
\end{equation}
and
\begin{equation}
\beta_{\vec{k}, +} = \frac{1}{\sqrt{2(1 + \frac{|k|^2 s^2}{4
\mu^2})}} [- \frac{|k| s}{2 \mu} ( \sqrt{\frac{k}{k^*}} a_{+} +
b_{+}) + (- \sqrt{\frac{k}{k^*}} a_{-}^{\dagger} +
b_{-}^{\dagger})],
\end{equation}
and negative eigenvalues,
\begin{equation}
\alpha_{\vec{k}, -} = \frac{1}{\sqrt{2(1 + \frac{4 \mu^2}{|k|^2
s^2})}} [\frac{2 \mu}{|k| s} ( \sqrt{\frac{k}{k^*}} a_{+} + b_{+}) +
(- \sqrt{\frac{k}{k^*}} a_{-}^{\dagger} + b_{-}^{\dagger})],
\end{equation}
and
\begin{equation}
\beta_{\vec{k}, -} = \frac{1}{\sqrt{2(1 + \frac{4 \mu^2}{|k|^2
s^2})}} [\frac{2 \mu}{|k| s} ( \sqrt{\frac{k}{k^*}} a_{+} - b_{+}) +
( \sqrt{\frac{k}{k^*}} a_{-}^{\dagger} + b_{-}^{\dagger})].
\end{equation}
 Similarly as before we can define
\begin{eqnarray}
\sqrt{\frac{k}{k^{*}}} a_{+} + b_{+} \equiv c_{+
 v},\\
\sqrt{\frac{k}{k^{*}}} a_{-}^{\dagger} + b_{-}^{\dagger}\equiv c_{- v}^{\dagger},\\
\sqrt{\frac{k}{k^{*}}} a_{+} - b_{+} \equiv c_{+ w},\\
-\sqrt{\frac{k}{k^{*}}} a_{-}^{\dagger} + b_{-}^{\dagger}\equiv c_{-
w}^{\dagger},
\end{eqnarray}
and the ground state vector can be cast in the following form,
\begin{equation}
\exp\{ \sum_{\vec{k}} \frac{2 \mu}{s |k|} ( c_{+,v}^{\dagger}
c_{-,w}^{\dagger} + c_{+,w}^{\dagger} c_{-,v}^{\dagger})\}|0\rangle.
\end{equation}
In this case each Cooper pair is antisymmetric under exchange of
$\vec{K}_{\pm}$ points i.e. valley degree of freedom and symmetric
under exchange of sublattices i.e. chirality $(v \leftrightarrow
w)$. Depending on our definitions for $c$'s two degrees of freedom
can exchange the symmetry properties. We find again the weak pairing
$(\sim \frac{1}{r})$ behavior in the orbital part.
\\
\\
\\
\section{Conclusions: Superconductivity on honeycomb lattice}
We derived the ground state wave functions for the superconductivity
on the honeycomb lattice induced by nearest-neighbor attractive
interactions and with order parameter independent of the direction
to any of the nearest neighbors. Although the order parameter in
momentum space has the $ p \pm i p$ form in a low effective
description the Cooper pair wave function behaves as $s$-wave (with
no angular dependence) and decays as $ \sim \frac{1}{r}$. Other
(discrete) degrees of freedom combine to make the Cooper pair
antisymmetric under exchange. At the point of the transition, $\mu =
0$, in the spin-singlet case a strong pairing (of the order of
lattice spacing) occurs.

\section{Introduction: Superconductivity on bilayer honeycomb lattice}
Topological superconductors in a strict sense or what we also call
non-trivial topological superconductors have odd number of Majorana
modes moving in each direction on the edge of such a superconductor
\cite{Qi}. In the case of trivial topological superconductors we
have even number of Majorana modes i.e. by combining them in pairs
we can talk about Dirac fermions on their edge. $s$-wave
superconductor in two dimensions is always topological in the sense
that it has a gap in its bulk and non-trivial degeneracy of the
ground state on the torus (equal to four) \cite{Hans}. We have to
use one Bose field (one Dirac fermion) to describe the edge of such
a system \cite{Hans}. On the other hand if we combine two $p$-wave
superconductors, with $p_x + i p_y$ and $p_x - i p_y$ orbital
symmetry, and each of the two corresponds to one let's say definite
projection of spin we have the case for a non-trivial topological
superconductor. In that case on the edge live two Majorana modes
that are moving in opposite directions and each is associated with
different projection of electron spin. This represents a ``helical"
edge where we have a pair of edge Majorana modes (moving in opposite
directions) that are connected with a time-reversal operation. This
is the simplest topological superconductor we can imagine in two
dimensions and has yet to be realized and detected in experiments.
In three dimensions a realization of topological superconductor is
He$^3$ B - phase \cite{Qi}.

On the other hand the honeycomb lattice, nowadays very much
connected with the research on graphene, is a playing ground for
various, among others topological, phases. The first topological
insulator was introduced on the honeycomb lattice with a special
interaction \cite{MeKa}. While considering possibilities for
superconducting instabilities on the honeycomb lattice and graphene
in Ref. \onlinecite{ucho}, a phase was proposed with two $p$-wave
order parameters (each near two effective descriptions in
$\vec{k}$-space i.e. two valleys). Though one might expect that,
while considering triplet pairing i.e. if we suppress spin, this
would lead to a non-trivial topological superconductor with a pair
of Majorana modes, this is not the case as we demonstrated in the
first part of the paper (Ref. \onlinecite{jrp}). We found that the
ground state wave function is antisymmetric with respect to the
valley degree of freedom and that the orbital part of the Cooper
pair is with no angular dependence i.e. a $s$-wave. As we emphasized
earlier in the case of $s$-wave we expect one Dirac fermion per
degree of freedom on the edge i.e. no  non-trivial behavior.

In this paper we will derive ground state wave functions for some
superconducting instabilities that may emerge due attractive
interactions on a two layer system in which each layer represents a
honeycomb lattice. The two lattices are stacked as in the bilayer
graphene i.e. in the way of Bernal stacking. As in the bilayer
graphene, we expect that the pseudospin vector connected with
sublattice degrees of freedom will not follow the momentum vector in
a parallel or antiparallel fashion, like on ordinary honeycomb
lattice or graphene, but rotate for a whole angle as a result of a
rotation of the momentum vector for a half an angle. This feature of
free electrons on the bilayer honeycomb lattice will reflect in the
description of Cooper pairs when attractive interactions are
introduced. The explicit ground state wave functions and Cooper pair
structure will help us to see more closely the nature of pairing in
this system. We will find the $p$-wave angular dependence in the
orbital part.

In the following section, we will formulate Bogoliubov - de Gennes
(BdG) equatons for this system. In the next section the explicit
solutions with corresponding ground state wave functions will be
given in the case of (a) spin-singlet and (b) spinless
(spin-triplet) pairing. Then we will examine whether these systems
are truly gapped in the bulk, and, in the spin-singlet case, its
edge spectrum. The last section is devoted to discussion and
conclusions.

\section{Electrons on bilayer honeycomb lattice and BCS instability}

\begin{figure}
\centering
\includegraphics[scale=0.4]{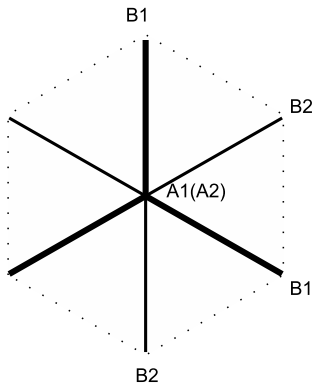}
\caption{A view of Bernal stacked honeycomb lattices 1 and 2 with
corresponding sublattice sites A1 and B1, and A2 and B2.}
\end{figure}

The Hamiltonian for free electrons on two honeycomb lattices, which
are Bernal stacked, is
\begin{equation}
H_{0} = - t \sum_{\vec{n},\sigma} \sum_{\vec{\delta}}
(a_{1,\vec{n},\sigma}^{\dagger} b_{1, \vec{n} + \vec{\delta},\sigma}
+ a_{2,\vec{n},\sigma}^{\dagger} b_{2, \vec{n} -\vec{\delta},\sigma}
+ h.c.) + t_{\bot} \sum_{\vec{n},\sigma}
(a_{1,\vec{n},\sigma}^{\dagger} a_{2, \vec{n},\sigma} + h.c) - \mu
\sum_{\vec{n}} \hat{n}_{\vec{n}}. \label{freeham}
\end{equation}
The index $i = 1,2$ denotes the layer index. In Fig. 1 the relative
positions of two triangular sublattices, $A_1$ and $B_1$, for the
lattice 1, and $A_2$ and $B_2$, for the lattice 2 are illustrated.
In Eq.(\ref{freeham}) $t$ is the hopping energy between nearest
neighbor C (carbon) atoms in the case of the bilayer graphene in
each layer, and $t_{\bot}$ is the same energy for hopping between
the layers. The on-site  creation (annihilation) operators,
$a_{i,\vec{n},\sigma}^{\dagger} (a_{i, \vec{n},\sigma})$, are for
the electrons in the sublattice $A_i$ of the layer $i$ with spin
$\sigma = \uparrow, \downarrow$, and $b_{i,\vec{n},\sigma}^{\dagger}
(b_{i, \vec{n},\sigma})$ for the electrons in the sublattice $B_i$,
$\hat{n}_{\vec{n}}$ is the on-site number operator, and $\mu$ is the
chemical potential. $\vec{\delta}$'s are defined as $\delta_{1} = a
(0, 1/\sqrt{3})$, $\delta_{2} = a/2 (1, - 1/\sqrt{3})$, and
$\delta_{3} = a/2 (- 1, - 1/\sqrt{3})$, and $ a = \sqrt{3}\;
a_{cc}$, $a_{cc}$ is the distance between $C$ atoms and $a$ is the
next to nearest neighbor distance.

We use units such that $\hbar = 1$. By introducing Fourier
transforms $ a_{i, \vec{k}, \sigma} = \sum_{\vec{n}} a_{i, \vec{n},
\sigma} \exp\{ i \vec{k} \; \vec{n}\}$ and $ b_{i, \vec{k}, \sigma}
= \sum_{\vec{n}} b_{i, \vec{n}, \sigma} \exp\{  i \vec{k} \;
\vec{n}\}$ etc. and diagonalizing the Hamiltonian we find for the
spectrum,
\begin{equation}
E^{\pm}_{\alpha} (\vec{k}) = \pm ((-1)^{\alpha} \frac{t_{\bot}}{2} +
\sqrt{\frac{t_{\bot}^{2}}{4} + t^2 |S(\vec{k})|^{2}})\;,
\end{equation}
where $S(\vec{k}) = \sum_{\vec{\delta}} \exp\{i \vec{k}\;
\vec{\delta}\}$, and $\alpha = 1,2$ stand for two kinds of branches.
Near $\vec{K}$ points, the corners of the hexagonal Brillouin zone,
$\vec{K}_{\pm} = (2 \pi)/a (\pm 2/3,0)$, we have
\begin{equation}
S(\vec{K}_{\pm} + \vec{k}) \approx \mp a \sqrt{3}/2 (k_{x} \mp i
k_{y}),
\end{equation}
and in the limit $t_{\bot} \gg t$ the lower positive and higher
negative branch have the folowing dispersion relation,
\begin{equation}
E_1^{\pm} = \pm \frac{\vec{k}^{2}}{2 m^{*}},
\end{equation}
where $ m^* = \frac{t_\bot}{2 v_B^2}$ and $ v_F = (\sqrt{3} a t)/2$,
the Fermi-Dirac velocity. The effective Hamiltonian near $\vec{K}$
points \cite{McFa} is
\begin{equation}
H_{ef}(\vec{k})= - \frac{v_F^2}{t_\bot} \left[\begin{array}{cc}  0 & (k^*)^2 \\
                                      (k)^2  & 0 \\

\end{array} \right],
\end{equation}
and acts on the subspace of (pseudo)spinors
\begin{equation}
\Psi_+ = [ b_{2,\sigma(\vec{K}_{+} + k)}, b_{1,\sigma(\vec{K}_{+} +
k)}]^{T}
\end{equation}
around point $\vec{K}_{+}$, and
\begin{equation}
\Psi_- = [ b_{1,\sigma(\vec{K}_{-} + k)}, b_{2,\sigma(\vec{K}_{-} +
k)}]^{T}
\end{equation}
around point $\vec{K}_{-}$. $H_{ef}$ can be rewritten as
\begin{equation}
H_{ef}(\vec{k}) = - \frac{1}{2 m^*} [(k_x^2 - k_y^2) \sigma_x + 2
k_x k_y \sigma_y] = -\frac{\vec{k}^2}{2 m^*} \vec{\sigma} \vec{n},
\end{equation}
where $\vec{k} = |\vec{k}|(\cos\{\phi_{\vec{k}}\},
\sin\{\phi_{\vec{k}}\})$ and $\vec{n} = (\cos\{2 \phi_{\vec{k}}\},
\sin\{2 \phi_{\vec{k}}\})$, and $\sigma$'s are Pauli matrices. The
operator $\vec{\sigma} \vec{n}$ encodes the projection of the
pseudospin on direction $\vec{n}$. For eigenstates as
\begin{equation}
\chi^+ = \left[\begin{array}{c}  \frac{k^{*}}{k} \\
                                   1
\end{array} \right],\label{sps} \;\; {\rm and}\;\;
\chi^- = \left[\begin{array}{c}  \frac{k^{*}}{k} \\
                                   -1
\end{array} \right],
\end{equation}
the direction $\vec{n}$ may be interpreted as the direction of the
pseudospin vector, with projection (chirality) equal to $+ 1$ in the
case of $\chi^+$, and $- 1$ in the case of $\chi^-$. Thus in the
case of these eigenstates we see explicitly our previous remark that
the pseudospin vector rotates for an angle while $\vec{k}$ vector
rotates for half an angle circling the Fermi surface around
$\vec{K}$ points. That feature of the solutions of the free problem
leads to non-trivial pairing in the orbital part of Cooper pairs as
we will see later. This is to be contrasted to the behavior in the
monolayer, a single honeycomb lattice, where the rotation of
$\vec{k}$ vector is strictly followed by the rotation of the
pseudospin vector. It is accompanied by $s$-wave pairing, when
special (nearest-neighbor) attractive interactions are applied. In
that case although two order parameters are of, $p_x + i p_y$, and
$p_x - i p_y$ type we have the trivial ($s$-wave) behavior in the
orbital part as we have shown earlier.

As the reader may have noticed we did not include the direct hopping
between the atoms of B1 and B2 sublattice. This inclusion is
required when we model bilayer graphene \cite{McFa}, but even there
for realistic parameters this does not influence the physics at high
electron momenta or strong magnetic fields \cite{novbi}.

But we will consider nearest-neighbor attractive interactions
between electrons on B1 and B2 sublattice. Namely these sublattices
by themselves make a honeycomb lattice as we can verify by looking
at Fig. 1.  Due to the strong hopping between A1 and A2 sublattice
the complete low-energy physics is projected onto B1 and B2
sublattice. If the interactions are not too strong they can be
simply added to this low-energy subspace. The on-site repulsive
interactions can be introduced and we do not expect that will change
our conclusions. Therefore the complete Hamiltonian will include
nearest-neighbor attractive interactions between electrons on B1 and
B2 sublattice as follows,
\begin{equation}
H_{I} = g \sum_{\vec{n},\vec{\delta}} \sum_{\sigma,\sigma'}
b_{1,\vec{n},\sigma}^{\dagger} b_{1, \vec{n},\sigma} b_{2,\vec{n} +
\vec{\delta},\sigma'}^{\dagger} b_{2, \vec{n} +
\vec{\delta},\sigma'},
\end{equation}
where $g < 0$. We will assume the spin-singlet pairing among nearest
neighbors and apply the BCS ansatz with
\begin{equation}
\Delta_{\vec{\delta}} = \langle b_{1,\vec{n},\uparrow} b_{2, \vec{n}
+ \vec{\delta},\downarrow} - b_{1,\vec{n},\downarrow} b_{2, \vec{n}
+ \vec{\delta},\uparrow} \rangle,
\end{equation}
the superconducting order parameter. Furthermore we assume one and
the same $\Delta_{\vec{\delta}} = \Delta$ for all nearest neighbors,
which due to global gauge (U(1)) transformations on $b_{1}$'s and
$b_{2}$'s can be chosen real and positive. The interaction part,
$H_I$, becomes
\begin{equation}
\tilde{H}_{BCS} = \{ g \sum_{\vec{n}, \vec{\delta}} \Delta (
b_{1,\vec{n}, \uparrow}^{\dagger} b_{2, \vec{n} + \vec{\delta},
\downarrow}^{\dagger} - b_{1,\vec{n}, \downarrow}^{\dagger} b_{2,
\vec{n} + \vec{\delta}, \uparrow}^{\dagger}) + h.c.\} - 3 g
|\Delta|^{2}.
\end{equation}
The order parameter in the momentum space is
\begin{equation}
\Delta_{\vec{k}}  =  \sum_{\vec{\delta}} \Delta \exp\{i \vec{k}
\vec{\delta}\} = \Delta S(\vec{k}).
\end{equation}
Therefore near $\vec{K}$ points $\Delta_{\vec{K}_{\pm} + \vec{k}}
\sim \mp (k_{x} \mp i k_{y})$, which then describes two $p$-wave
like superconducting order parameters in a low-energy effective
description. Taking into account the complete low-energy reduction
the total BCS Hamiltonian can now be cast in the following form in
the momentum space near $\vec{K}_{+}$, $\vec{q} = \vec{K}_{+} +
\vec{k}$,
\begin{equation}
 H_{BCS} = \sum_{\vec{q}} \phi_{\vec{q}}^{\dagger} M_{\vec{q}}
 \phi_{\vec{q}}, \label{hbcs2}
 \end{equation}
 where
 \begin{equation}
 \phi_{\vec{q}}^{\dagger} = (b_{2,\vec{q}, \uparrow}^{\dagger}, b_{1,\vec{q}, \uparrow}^{\dagger}
b_{2, -\vec{q}, \downarrow}, b_{1, -\vec{q}, \downarrow})
\end{equation}
 and, with $ s = s^* \equiv - \Delta g a \sqrt{3}/2 > 0$,
\begin{displaymath}
M_{\vec{q}}|_{\vec{q} = \vec{K}_{+} +
\vec{k}}\equiv M_{\vec{k}} = \left[\begin{array}{cccc} - \mu & - \frac{(k^*)^2}{2 m^*} & 0 & s k^* \\
                                   - \frac{(k)^2}{2 m^*} & - \mu & s k  & 0 \\
                          0 & s k^* & \mu &   \frac{(k^*)^2}{2 m^*}  \\
                          s k & 0 &  \frac{(k)^2}{2 m^*}  &
                          \mu
\end{array} \right].
\end{displaymath}
We will omit the discussion concerning $M_{\vec{q}}$ in the
neighborhood of $\vec{K}_{-}$ and momenta: $\vec{q} = \vec{K}_{-} -
\vec{k}$. This entails operators which combine $\uparrow$ spin with
momenta $\vec{q} = \vec{K}_{-} - \vec{k}$ and $\downarrow$ spin with
momenta $\vec{q} = \vec{K}_{+} + \vec{k}$, and will not provide any
new information for the structure of the ground state wave function
or energy dispersion at small momenta. We can simply include these
operators at the end in the ground state wave function following
symmetry requirements for the spin-singlet pairing.

\section{Ground state wave functions of superconducting instabilities}
We look for the solution of Eq. (\ref{hbcs2}) in the form of a
diagonalized Bogoliubov BCS Hamiltonian,
\begin{equation}
H_{BCS} = \sum_{\vec{k}, \gamma = \pm} \omega_{\vec{k},
\gamma}^{\alpha} \alpha^{\dagger}_{\vec{k}, \gamma} \alpha_{\vec{k},
\gamma} + \sum_{\vec{k}, \gamma = \pm} \omega_{\vec{k},
\gamma}^{\beta} \beta^{\dagger}_{\vec{k}, \gamma}  \beta_{\vec{k},
\gamma} + E_{0}\;,
\end{equation}
where $\alpha_{\vec{k}, \gamma}$ and $\beta_{\vec{k}, \gamma}$,
$\gamma =\pm $ are new quasiparticles at momentum $\vec{k}$. For the
dispersions we have:
\begin{equation}
\omega_{\vec{k}, \gamma}^{\alpha} = \gamma
\omega_{\vec{k}}^{\alpha}\;\;\; {\rm and}\;\;\; \omega_{\vec{k},
\gamma}^{\beta} = \gamma \omega_{\vec{k}}^{\beta}\;,
\end{equation}
where $\gamma = \pm $. We define a general solution $\alpha$ as
\begin{equation}
\alpha_{\vec{k}} = u_{\vec{k}, \uparrow} b_{2,\vec{k}, \uparrow} +
v_{\vec{k}, \uparrow} b_{1, \vec{k}, \uparrow} + u_{\vec{k},
\downarrow} b_{2, - \vec{k}, \downarrow}^{\dagger} + v_{\vec{k},
\downarrow} b_{1, - \vec{k}, \downarrow}^{\dagger}.\label{gensol}
\end{equation}
Next we have to solve the Bogoliubov - de Gennes (BdG) equations,
which follow from the following condition,
\begin{equation}
 [\alpha_{\vec{k}}, H_{BCS}] = E \alpha_{\vec{k}}.\label{beq2}
\end{equation}
We need to diagonalize the following matrix, $M^{*}_{\vec{k}}$, that
comes out of Eq. (\ref{beq2}):
\begin{equation}
\left[\begin{array}{cccc}- \mu & - \frac{(k)^2}{2 m^*} & 0 & s k \\
                                   - \frac{(k^*)^2}{2 m^*} & - \mu & s k^*  & 0 \\
                          0 & s k & \mu &   \frac{(k)^2}{2 m^*}  \\
                          s k^* & 0 &  \frac{(k^*)^2}{2 m^*}  &
                          \mu
\end{array} \right].
\end{equation}
From this matrix eigenvalue problem we obtain energies of the
Bogoliubov quasiparticles,
\begin{equation}
E_{\vec{k}}^{p} = \pm \sqrt{\mu^2 + |k|^2 s^2 + (\frac{1}{2 m^*})^2
|k|^4 + p \sqrt{4 (\frac{\mu}{2 m^*})^2 |k|^4 + 2 (\frac{s}{2
m^*})^2 |k|^6 - (k^*)^2 |k|^4 (\frac{s}{2 m^*})^2 - (k)^2 |k|^4
(\frac{s}{2 m^*})^2}}, \label{E1}
\end{equation}
where $\pm$ stands for the particle and hole branches respectively
for two kinds of excitations $p = - 1 (\alpha)$ and $ p = +
1(\beta)$. The eigenvectors of matrix $M^{*}_{\vec{k}}$ (after
normalization) enter the following expressions in the long-distance
limit for Bogoliubov quasiparticles :
\begin{equation}
\alpha_{\vec{k},+} = \frac{1}{\sqrt{2}} \{(\frac{s k}{2 \mu}) (b_{2
+ \uparrow} + b_{1 + \uparrow}) + (\frac{k}{k^{*}} b_{2 -
\downarrow}^{\dagger} + b_{1 - \downarrow}^{\dagger})\},
\end{equation}
and
\begin{equation}
\beta_{\vec{k},+} =  \frac{1}{\sqrt{2}} \{(\frac{s k}{2 \mu}) (b_{2
+ \uparrow} - b_{1 + \uparrow}) - (\frac{k}{k^{*}} b_{2 -
\downarrow}^{\dagger} - b_{1 - \downarrow}^{\dagger})\},
\label{beta1}
\end{equation}
and quasiholes:
\begin{equation}
\alpha_{\vec{k},-} = \frac{1}{\sqrt{2}} \{ -(\frac{k}{k^{*}}b_{2 +
\uparrow} + b_{1 + \uparrow}) +  (\frac{s k}{2 \mu})(b_{2 -
\downarrow}^{\dagger} + b_{1 - \downarrow}^{\dagger})\} ,
\end{equation}
and
\begin{equation}
\beta_{\vec{k},-} = \frac{1}{\sqrt{2}} \{ -(\frac{k}{k^{*}}b_{2 +
\uparrow} - b_{1 + \uparrow}) -  (\frac{s k}{2 \mu})(b_{2 -
\downarrow}^{\dagger} - b_{1 - \downarrow}^{\dagger})\}
,\label{beta2}
\end{equation}
for the Bogoliubov solution near point $\vec{K}_+$, where we denoted
$b_{2,\vec{K}_\pm \pm \vec{k}, \sigma} \equiv b_{2 \pm \sigma}$ and
$b_{1,\vec{K}_\pm \pm \vec{k}, \sigma} \equiv b_{1 \pm \sigma}$.

It is helpful to introduce the creation and annihilation operators
of states of definite chirality or projection of the pseudospin
along the $\vec{n}$ vector; we denote by $v$ positive projection,
and by $w$ negative projection:
\begin{eqnarray}
&&c_{+ v} = \frac{k}{k^*} b_{2 + \uparrow} + b_{1 + \uparrow}, \\
&&c_{+ w} = \frac{k}{k^*} b_{2 + \uparrow} - b_{1 + \uparrow}, \\
&&c_{- v} = \frac{k^*}{k} b_{2 - \downarrow} + b_{1 - \downarrow}, \\
&&c_{- w} = - \frac{k^*}{k} b_{2 - \downarrow} + b_{1 - \downarrow}.
\end{eqnarray}
Notice that the states are defined up to a phase factor according to
the definitions in Eq.(\ref{sps}), and that in the states around
$\vec{K}_{-}$ point the roles of electrons in different layers are
interchanged. We may then define their superpositions,
\begin{eqnarray}
&&c_{\pm r} = (c_{\pm v} + c_{\pm w})/2,\\
&&c_{\pm l} = (c_{\pm v} - c_{\pm w})/2.
\end{eqnarray}
Then we can rewrite Bogoliubov operators as:
\begin{eqnarray}
&&\alpha_{k, +} = \frac{1}{\sqrt{2}}\{r k c_{+ l} + r k^*
c_{+ r} + c_{- r}^{\dagger} + c_{- l}^{\dagger}\}, \\
&&\beta_{k, +} = \frac{1}{\sqrt{2}}\{- r k c_{+ l} + r k^* c_{+ r} +
c_{- r}^{\dagger} - c_{- l}^{\dagger}\},\\
 &&\alpha_{k, -} = \frac{1}{\sqrt{2}}\{r k c_{- r}^{\dagger} + r k^*
c_{- l}^{\dagger} - c_{+ r} - c_{+ l}\}, \\
&&\beta_{k, -} = \frac{1}{\sqrt{2}}\{r k c_{- r}^{\dagger} - r k^*
c_{- l}^{\dagger} - c_{+ r} + c_{+ l}\},
\end{eqnarray}
with $r \equiv \frac{s}{2 \mu}$, and find that the wave function,
$\tilde{\Psi}_{0}$,
\begin{equation}
\tilde{\Psi}_{0} = (1 - \frac{1}{r k^*} c_{+ r}^{\dagger} c_{-
r}^{\dagger} - \frac{1}{r k} c_{+ l}^{\dagger} c_{- l}^{\dagger} +
\frac{1}{r^2 |k|^2} c_{+ r}^{\dagger} c_{- r}^{\dagger} c_{+
l}^{\dagger} c_{- l}^{\dagger})|0\rangle
\end{equation}
is annihilated by quasiparticle annihilation operators, $ \alpha_{k,
+} \tilde{\Psi}_{0} = \beta_{k, +} \tilde{\Psi}_{0} = 0$, and by
quasihole creation operators, $ \alpha_{k, -}^{\dagger}
\tilde{\Psi}_{0} = \beta_{k, -}^{\dagger} \tilde{\Psi}_{0} = 0$.
Therefore the new ground state is
\begin{equation}
\Psi_{0} = \exp\{ - \sum_{k} \frac{1}{r k^*} (c_{+ r
\uparrow}^{\dagger} c_{- r \downarrow}^{\dagger} - c_{+ r
\downarrow}^{\dagger} c_{- r \uparrow}^{\dagger}) - \sum_{k}
\frac{1}{r k} (c_{+ l \uparrow}^{\dagger} c_{- l
\downarrow}^{\dagger} - c_{+ l \downarrow}^{\dagger} c_{- l
\uparrow}^{\dagger})\}|0\rangle, \label{gs1}
\end{equation}
where we explicitly introduced the sector that couples momenta
around $\vec{K}_{+}$ with spin $\downarrow$ and momenta around
$\vec{K}_{-}$ with spin $\uparrow$ by enforcing the explicit
spin-singlet pairing which we introduced at the beginning.

From the structure of the ground state wave function for
spin-singlet pairing in Eq.(\ref{gs1}) we find that electrons pair
between $\vec{K}_{+}$ and $\vec{K}_{-}$ point have the same
pseudospin and therefore we have two distinct Cooper pairings for
two orthogonal pseudospin states, which we denoted by $r$ and $l$.
Each pair has a $p$-wave pairing in the orbital part and, as we work
with a time-reversal invariant system, two distinct Cooper pairings
are accompanied by two distinct, $p_x + i p_y$ and $p_x - i p_y$,
symmetries in the orbital part. Each Cooper pair is antisymmetric
under spin exchange, valley exchange, and exchange in the orbital
part and symmetric under sublattice (pseudospin) exchange. Therefore
the ground state wave function in Eq.(\ref{gs1}) describes a
Cooper-paired collection of fermions -electrons on the bilayer
honeycomb lattice with unconventional $p$-wave pairing.

The non-trivial (non-$s$-wave) pairing cannot be eliminated with a
gauge transformation. To preserve the form of the free part of the
Hamiltonian any gauge transformation should be $ b_{2 +
\sigma}^{\dagger} \rightarrow \exp\{i \phi_{+}\} b_{2 +
\sigma}^{\dagger}$ and $ b_{1 + \sigma}^{\dagger} \rightarrow
\exp\{i \phi_{+}\} b_{1 + \sigma}^{\dagger}$, and $ b_{2 -
\sigma}^{\dagger} \rightarrow \exp\{i \phi_{-}\} b_{2 -
\sigma}^{\dagger}$ and $ b_{1 - \sigma}^{\dagger} \rightarrow
\exp\{i \phi_{-}\} b_{1 - \sigma}^{\dagger}$, and if we rewrite
$\Psi_{0}$ in terms of these operators
\begin{equation}
\Psi_{0} = \exp\{ - \sum_{k} \frac{1}{r k} (b_{2 +
\uparrow}^{\dagger} b_{1 -  \downarrow}^{\dagger} - b_{2 +
\downarrow}^{\dagger} b_{1 - \uparrow}^{\dagger}) - \sum_{k}
\frac{1}{r k^*} (b_{1 +  \uparrow}^{\dagger} b_{2 -
\downarrow}^{\dagger} - b_{1 +  \downarrow}^{\dagger} b_{2 -
\uparrow}^{\dagger})\}|0\rangle, \label{gs11}
\end{equation}
we see that by this gauge transformation we cannot eliminate
simultaneously the angular dependence in the two types of Cooper
$p$-wave pairings.

Next we will discuss the spinless case. We will assume that all
electron spins are polarized and that $ \langle b_{1 \vec{n}} b_{2
\vec{n} + \vec{\delta}}\rangle = \Delta$. In this case the
Bogoliubov problem in Eq.(\ref{hbcs2}) for the spin-singlet pairing
transforms into a similar one with $ b_{1 \vec{k} \sigma} \equiv
b_{1 \vec{k}}$ and $ b_{2 \vec{k} \sigma} \equiv b_{2 \vec{k}}$ and
the matrix $M_{\vec{q}}$ becomes as follows,
\begin{displaymath}
M_{\vec{q}}|_{\vec{q} = \vec{K}_{+} +
\vec{k}}\equiv M_{\vec{k}} = \left[\begin{array}{cccc} - \mu & - \frac{(k^*)^2}{2 m^*} & 0 & - s k^* \\
                                   - \frac{(k)^2}{2 m^*} & - \mu & s k  & 0 \\
                          0 & s k^* & \mu &   \frac{(k^*)^2}{2 m^*}  \\
                         - s k & 0 &  \frac{(k)^2}{2 m^*}  &
                          \mu
\end{array} \right].
\end{displaymath}
The problem around the $\vec{K}_{-}$ point is a copy of the problem
around $\vec{K}_{+}$. We find the following eigenvalues for the
Eq.(\ref{beq2}),
\begin{equation}
E_{\vec{k}}^{p} = \pm \sqrt{\mu^2 + |k|^2 s^2 + (\frac{1}{2 m^*})^2
|k|^4 + p \sqrt{4 (\frac{\mu}{2 m^*})^2 |k|^4 + 2 (\frac{s}{2
m^*})^2 |k|^6 + (k^*)^2 |k|^4 (\frac{s}{2 m^*})^2 + (k)^2 |k|^4
(\frac{s}{2 m^*})^2}}, \label{E2}
\end{equation}
where $\pm$ stands for the particle and hole branches respectively
for two kinds of excitations $p = - 1 (\alpha)$ and $ p = +
1(\beta)$. The eigenvectors  enter the following expressions in the
long-distance limit for Bogoliubov quasiparticles :
\begin{equation}
\alpha_{\vec{k},+} = \frac{1}{\sqrt{2}} \{(\frac{s k}{2 \mu}) (-
b_{2 +} + b_{1 +}) + (\frac{k}{k^{*}} b_{2 -}^{\dagger} + b_{1
-}^{\dagger})\},
\end{equation}
and
\begin{equation}
\beta_{\vec{k},+} =  \frac{1}{\sqrt{2}} \{(\frac{s k}{2 \mu}) (-
b_{2 +} - b_{1 +}) - (\frac{k}{k^{*}} b_{2 -}^{\dagger} - b_{1
-}^{\dagger})\},
\end{equation}
and quasiholes:
\begin{equation}
\alpha_{\vec{k},-} = \frac{1}{\sqrt{2}} \{ (\frac{k}{k^{*}}b_{2 +} +
b_{1 +}) - (\frac{s k}{2 \mu})(b_{2 -}^{\dagger} - b_{1
-}^{\dagger})\} ,
\end{equation}
and
\begin{equation}
\beta_{\vec{k},-} = \frac{1}{\sqrt{2}} \{ (\frac{k}{k^{*}}b_{2 +} -
b_{1 +}) +  (\frac{s k}{2 \mu})(b_{2 -}^{\dagger} + b_{1
-}^{\dagger})\}.
\end{equation}
Introducing as in the spin-singlet case the following pseudospin
operators:
\begin{eqnarray}
&c_{- l}^{\dagger} = \frac{k}{k^*} b_{2 -}^{\dagger}\;\; &{\rm
and}\;\; c_{- r}^{\dagger} = b_{1 -}^{\dagger}\\
&c_{+ r}^{\dagger} = \frac{k^*}{k} b_{2 +}^{\dagger}\;\; &{\rm
and}\;\; c_{+ l}^{\dagger} = b_{1 +}^{\dagger}
\end{eqnarray}
we can rewrite the eigenvectors as follows,
\begin{eqnarray}
&&\alpha_{k, +} = \frac{1}{\sqrt{2}}\{r k c_{+ l} - r k^*
c_{+ r} + c_{- r}^{\dagger} + c_{- l}^{\dagger}\}, \\
&&\beta_{k, +} = \frac{1}{\sqrt{2}}\{- r k c_{+ l} - r k^* c_{+ r} +
c_{- r}^{\dagger} - c_{- l}^{\dagger}\},\\
 &&\alpha_{k, -} = \frac{1}{\sqrt{2}}\{r k c_{- r}^{\dagger} - r k^*
c_{- l}^{\dagger} + c_{+ r} + c_{+ l}\}, \\
&&\beta_{k, -} = \frac{1}{\sqrt{2}}\{r k c_{- r}^{\dagger} + r k^*
c_{- l}^{\dagger} + c_{+ r} - c_{+ l}\}.
\end{eqnarray}
Similarly as in the spin-singlet case we can find that the ground
state wave function can be expressed as
\begin{eqnarray}
\Psi_{0}& = & \exp\{\sum_{k} \frac{1}{r k^*} c_{+ r}^{\dagger} c_{-
r}^{\dagger} - \sum_{k} \frac{1}{r k} c_{+ l}^{\dagger} c_{-
l}^{\dagger}\}|0\rangle\label{gs2} \\
& = & \exp\{\sum_{k} \frac{1}{r k} b_{2 +}^{\dagger} b_{1
-}^{\dagger} - \sum_{k} \frac{1}{r k^*} b_{2 -}^{\dagger} b_{1
+}^{\dagger}\}|0\rangle.
\end{eqnarray}
The Cooper pairs are symmetric under valley and sublattice
(pseudospin) exchange and antisymmetric under exchange in the
orbital part. We have two kinds of Cooper pairs with underlying $p_x
+ i p_y$ and $p_x - i p_y$ symmetry.

\section{The nature of pairing phases}

\subsection{Gaps and possible nodes}

In Ref. \onlinecite{blh} the case of spin-singlet pairing in the
monolayer was thoroughly discussed. A topological phase structure
was described with four (due to valley and spin) Dirac edge modes.
This is consistent with our previous calculations of the ground
state wave function that has explicit $s$-wave dependence $\sim
\frac{1}{|k|}$ for $ |\mu| > 0$ (chemical potential). Here we
extended ground state  calculations to the spin-singlet and
spin-triplet case of the bilayer. It is appropriate to ask the
question raised in Ref. \onlinecite{blh} for spin-singlet and
spin-triplet monolayer case: Are these phases  truly gapped or there
are nodes for some $k$'s in the bulk spectrum ? In the same
reference it was found that in the spin-triplet case, as opposed to
the spin-singlet case, there are nodes in $k$ space at which gap is
equal to zero. We will find a similar situation in the bilayer case,
except that in the case of spin-singlet pairing there is a critical
value for chemical potential above which we have a truly gapped -
topological phase. (The triplet case just as in the monolayer
analysis in Ref. \onlinecite{blh} has nodes in the bulk spectrum.)

It is not hard, by repeating the approach of Ref.\onlinecite{McFa},
to find expressions for the matrix, $M_{\vec{k}}^{*}$, that enters
BdG equations in both cases for general (not low) momentum $k$. They
are
\begin{equation}
M_{\vec{k}}^{*} = \left[\begin{array}{cccc} - \mu & - T (S^*(\vec{k}))^2 & 0 & \Delta S^*(\vec{k}) \\
                                    - T (S(\vec{k}))^2& - \mu & \pm \Delta S(\vec{k})  & 0 \\
                          0 & \pm \Delta S^{*}(\vec{k}) & \mu &  T (S^*(\vec{k}))^2  \\
                          \Delta S(\vec{k}) & 0 & T (S(\vec{k}))^2  &
                          \mu
\end{array} \right]\label{mat},
\end{equation}
where $ T \equiv \frac{t^2}{t_\bot}$ and $+$ and $-$ stand for
spin-singlet and spin-triplet case respectively. We look for the
zeros of Bogoliubov quasiparticle energies, expressed in the low $k$
limit in Eq.(\ref{E1}) and Eq.(\ref{E2}), when $p = - 1$. With no
low momentum limit, from Eq.(\ref{mat}), their expressions are,
\begin{equation}
E^{\mp} = \sqrt{\mu^2 + |S(k)|^2 \Delta^2 + |S(k)|^4 v^2 - \sqrt{4
|S(k)|^4 \mu^2 T^2 + \Delta^2 T^2 |S(k)|^4 [2 |S(k)|^2 \mp S(k)^2
\mp (S^*(k)^2)]}},
\end{equation}
where we used shorthand notation taking $ g \Delta \equiv \Delta$,
and $E^{-}$ and $E^{+}$ correspond to the spin-singlet and
spin-triplet case respectively. We assume that $\Delta$ is small
with respect to $\mu$ so that possible nodes can be only near Fermi
surface defined by $|\mu| = T |S(k)|^2$. Equating $E^{-}$ and
$E^{+}$ to zero is equivalent to the following condition,
\begin{equation}
(\mu^2 - T^2 |S|^4)^2 = - |S|^4 \Delta^4 - 2 \mu^2 |S|^2 \Delta^2
\mp 2 \Delta^2 T^2 |S|^6 \cos\{2 \phi\},
\end{equation}
where $\phi$ is the phase of $S(k)$ - complex number in general. If
we assume that we work with $k$'s near Fermi surface we have
approximately
\begin{equation}
2 \mu (1 \pm \cos\{2 \phi\}) \approx \mp 4 T cos\{2 \phi\} \delta -
\frac{1}{T} \Delta^2, \label{cond}
\end{equation}
where $\delta$ is defined by $|s|^2 = \frac{\mu}{T} + \delta$ as a
small depature from the Fermi surface value. Therefore we can
approximate that for possible nodes near Fermi surface in the case
of spin-singlet pairing $S(k)$ is imaginary i.e. $\phi = \pm
\frac{\pi}{2}$ and in the case of spin-triplet pairing $S(k)$ is
real i.e. $\phi = 0, \pi$.

It is not hard to find nodes in the spin-triplet case. From the
definition,
\begin{equation}
S(k) = \exp\{i \frac{k_y}{\sqrt{3}}\} + \exp\{i \frac{1}{2}(k_x -
\frac{k_y}{\sqrt{3}})\}  + \exp\{-i\frac{1}{2}(k_x +
\frac{k_y}{\sqrt{3}})\}, \label{def}
\end{equation}
we can recognize that possible positions of nodes can be restricted
to $k_x = \frac{k_y}{\sqrt{3}}$, because in that case $S(k)$ takes
real values; $S(k) = 1 + 2 \cos\{\frac{k_y}{\sqrt{3}}\}$. Then in
our approximation from Eq.(\ref{cond}) and definition $|S|^2 =
\frac{\mu}{T} + \delta$ we have
\begin{equation}
k_y = \sqrt{3} \arccos\{\frac{S(k)}{2} - \frac{1}{2}\} \approx
\sqrt{3} \arccos\{\frac{\sqrt{\frac{\mu}{T}} (1 + \frac{\Delta^2}{2
\mu T})}{2} - \frac{1}{2}\},
\end{equation}
and that with $k_x = \frac{k_y}{\sqrt{3}}$ defines a position of a
single node. Therefore the existence of this node (and other related
by symmetry) tell us that this phase is likely to be gapless even in
the bulk and can not represent a topological phase.

In the spin-singlet case, if we rescale the momentum $k_y$ as
$\frac{k_y}{\sqrt{3}} \rightarrow k_y$ in Eq.(\ref{def}), the
condition that $S(k)$ is purely imaginary demands that
\begin{equation}
\cos\{k_y\} + 2 \cos\{\frac{k_x}{2}\} \cos\{\frac{k_y}{2}\} = 0.
\end{equation}
Then
\begin{equation}
2 Im S(k) = \sin\{k_y\} - 2 \cos\{\frac{k_x}{2}\}
\sin\{\frac{k_y}{2}\} = \frac{\sin\{\frac{3
k_y}{2}\}}{\cos\{\frac{k_y}{2}\}}.
\end{equation}
Expressed differently as
\begin{equation}
(1 - \cos\{3 k_y\}) = 4 |S|^2 (1 + \cos\{k_y\}),
\end{equation}
this leads to the conclusion that for large enough $|S|^2$ i.e.
chemical potential this equation does not have a solution for
$\cos\{k_y\}$. We find that for $|S|^2 > \frac{3}{4} (\sqrt{12} - 3)
\approx 0.348$ no solution exists. Therefore for large enough
chemical potential we can have a spin-singlet topological phase i.e.
a phase with no gapless bulk excitations.
\subsection{Edge modes}
To further examine the topological nature of the spin-singlet phase
we will derive its edge modes. With respect to the lattice structure
we will consider a particular geometry where the system, defined on
a half-plane, has the edge at $x = 0$. We remind the reader that we
use the convention in which $\vec{K}_{\pm}$ vectors are along $x$
axis. This choice of boundary corresponds to so-called armchair
boundary condition for which we require that the solutions of BdG
equations vanish at $x = 0$.

We will consider BdG equations in the low $k$ limit around
$\vec{K}_{\pm}$ points, neglect terms quadratic in $k (k^*)$, and
employ the substitution $k_x \rightarrow - i
\frac{\partial}{\partial x}$ to get their form in the real space;
due to the symmetry of the problem we seek solutions in the form
$\sim \exp\{i k_y y\} f(x)$ and keep the $k_y$ dependence. The
expression for BdG matrix at momentum $\vec{K}_{+} + \vec{k}$ is
\begin{equation}
M_{\vec{q}}|^*_{\vec{q} = \vec{K}_{+} +
\vec{k}} = \left[\begin{array}{cccc} - \mu & 0 & 0 & s k \\
                                   0 & - \mu & s k^*  & 0 \\
                          0 & s k & \mu &  0  \\
                         s k^* & 0 &  0  &
                          \mu
\end{array} \right],
\end{equation}
and at momentum $\vec{K}_{-} - \vec{k}$,
\begin{equation}
M_{\vec{q}}|^*_{\vec{q} = \vec{K}_{-} -
\vec{k}} = \left[\begin{array}{cccc} - \mu & 0 & 0 & s k^* \\
                                   0 & - \mu & s k  & 0 \\
                          0 & s k^* & \mu &  0  \\
                         s k & 0 &  0  &
                          \mu
\end{array} \right],
\end{equation}
and we immediately see a reduction of the problem to four sets of
equations with the substitution $k_x \rightarrow - i
\frac{\partial}{\partial x}$. If we denote by
\begin{equation}
f(x) = (u_{\uparrow}, u_{\downarrow}, v_{\downarrow},
v_{\uparrow})^T,
\end{equation}
a form of a solution, where we used $\uparrow$ and $\downarrow$ to
denote in a shorthand notation {\em pseudospin} (compare with
Eq.(\ref{gensol}) where the same symbols were used for real spin),
we have four sets of  equations of similar form. For example, one
set at $K_+$ point is:
\begin{eqnarray}
 - \mu u_{\uparrow} - i s \frac{\partial}{\partial x} v_{\uparrow}
+
i s k_y v_{\uparrow} &=& E u_{\uparrow}\nonumber \\
\mu v_{\uparrow} - i s \frac{\partial}{\partial x} u_{\uparrow} -
 i s k_y u_{\uparrow} &=& E v_{\uparrow}.
\end{eqnarray}
Just as argued in Ref. \onlinecite{rg} in the case of a simple
$p$-wave superconductor, when $E = 0$ and $k_y = 0$ we have a zero
mode, $ u_{\uparrow} = i v_{\uparrow} \sim \exp\{- \frac{\mu}{s} x\}
\exp\{ i \frac{\pi}{4}\}$. The difference here is that $
u_{\uparrow}$ and $ v_{\uparrow}$ carry opposite valley indexes  and
we cannot construct Majorana Bogoliubov  quasiparticle with them,
but only a Dirac one. For $ E \neq 0$ we find a chiral
(unidirectional) mode: $E = s k_y, k_y > 0$. Considering analogous
equation for $u_{\uparrow}$ and $ v_{\uparrow}$ at $\vec{K}_{-}$
point we get an additional solution with the same chirality (the
same direction of $k_y$), but these two solutions enter the boundary
condition which requires that $ u_{\uparrow}$ vanishes at the
boundary. Therefore one solution with $ k_y > 0 $ is
\begin{equation}
 u_{\uparrow} \sim \exp\{i k_y y\} \exp\{-
\frac{\mu}{s} x\} \exp\{ i \frac{\pi}{4}\} \sin\{Q x\},\;\;
v_{\uparrow} = - i u_{\uparrow}, \;\; u_{\downarrow} =
v_{\downarrow} = 0,
\end{equation}
where $ Q = |\vec{K}_{\pm}|$. The other, with $ k_y < 0$ and $ E = -
s k_y$, we find in analogous way. Therefore, we find, as expected
from the ground state wave function in Eq.(\ref{gs1}), two (Dirac)
solutions with opposite directions of motion along the edge that
carry opposite pseudospin. In addition, due to the spin degree of
freedom we expect doubling of modes i.e. in total two Dirac modes in
each direction.
\section{Discussion and conclusions: Superconductivity on bilayer honeycomb lattice}
We do not have to go into a discussion of topological invariants to
recognize that ground states for spin-singlet pairing, Eq.
(\ref{gs1}), and for spin-triplet pairing, Eq.(\ref{gs2}), may
represent ground states of trivial topological superconductors; spin
and valley degree of freedom induce the doubling of Majorana modes
that may follow \cite{rg} from $p$-wave pairing in the orbital part.
Nevertheless these models of superconductors are interesting in
their own right due to the presence of the $p$-wave pairings in the
orbital part. In the case of spin-singlet pairing we found that
there are no gapless bulk excitations for large enough chemical
potential and this phase can represent a trivial topological
superconductor.

In deriving ground state wave functions, in the spin-singlet and
spin-triplet case, we assumed validity of small $k$ expansion around
$k = 0$. In a strict sense this requires that $k_F$ is in the same
neighborhood where we can approximate dispersion relations i.e.
either effective hopping (Fermi velocity) is large or chemical
potential is small. We also allowed the possibility that minima or
even nodes (in the spin-triplet case) in the spectra of the
superconductors can be away from $k = 0$, around $k_F$. In the cases
of the trivial topological superconductors, on the monolayer
 and bilayer honeycomb lattice, this seems completely
justified in the view of their edge spectrum (see also \cite{blh}).
In the cases of gapless spin-triplet superconductors derived ground
state wave functions, though we are inclined to associate them with
topological phases, may describe these critical states
\cite{comment}. We find that only if we apply bias to the bilayer
larger than its chemical potential we can not apply the BdG program
around $k = 0$ in the way we described in the paper. (We remind the
reader that with the bias the energy minima of the noninteracting
problem shift from $k = 0$ at $\vec{K}$ points to nonzero $k$'s.)

The attractive interactions that we need for the realization of the
paired ground states and phases may well be within the reach of
future experiments. In the case of a single honeycomb lattice the
interactions may be induced by chemically doping the graphene via
metal coating \cite{ucho} or trapping fermionic atoms in a honeycomb
optical lattice \cite{zhu}.

To conclude, in the second part of this paper we derived ground
state wave functions for the superconductivity on the bilayer
honeycomb lattice (with strong interlayer coupling) induced by
attractive interactions between sites that participate in a
low-energy description. As is well-known, without these
interactions, free electrons are described by a Dirac equation with
a quadratic dispersion. This unusual feature, similarly to $^{3}$He
- B phase, leads to the description with two kinds of Cooper pairs,
with $p_x + i p_y$ and $p_x - i p_y$ pairing, in the presence of the
attractive interactions. This is expressed in Eq.(\ref{gs1}) in the
case of the spin-singlet pairing. Due to the spin degree of freedom
we find doubling of two chiral Dirac modes with opposite pseudospin
on the edge of this spin-singlet superconductor - a trivial
topological superconductor.
\section{Acknowledgment}
This work was supported in part by the Ministry of Science and
Technological Development of the Republic of Serbia, under project
No. ON171017.

\end{document}